\documentclass[11pt]{extarticle}
\usepackage[paperheight=12in,paperwidth=8.75in]{geometry}
\usepackage{setspace}
\usepackage{amssymb,amsmath}
\usepackage{enumerate}
\usepackage{setspace}
\usepackage[T1]{fontenc}
\usepackage{times}
\usepackage{palatino}
\usepackage{lmodern}
\usepackage[latin1]{inputenc}
\usepackage{epsfig}
\usepackage[english]{babel}
\usepackage{color}
\usepackage{graphicx}
\usepackage{dcolumn}
\usepackage{amsmath,amssymb,amsfonts}
\usepackage[all]{xy}
\usepackage{cite}
\usepackage[dvipsnames]{xcolor}
\usepackage{mathrsfs}
\usepackage{authblk}
\usepackage{geometry}
\usepackage{abstract}
\usepackage[english]{babel}
\usepackage[T1]{fontenc}
\usepackage[colorlinks,urlcolor=blue,citecolor=blue]{hyperref}

\newtheorem{theorem}{Theorem}
\newcommand{\R}{\mathbb{R}}

\begin{document}
\numberwithin{equation}{section}
\newcommand{\boxedeqn}[1]{%
  \[\fbox{%
      \addtolength{\linewidth}{-2\fboxsep}%
      \addtolength{\linewidth}{-2\fboxrule}%
      \begin{minipage}{\linewidth}%
      \begin{equation}#1\end{equation}%
      \end{minipage}%
    }\]%
}

%\boxedeqn{}

\newsavebox{\fmbox}
\newenvironment{fmpage}[1]
     {\begin{lrbox}{\fmbox}\begin{minipage}{#1}}
     {\end{minipage}\end{lrbox}\fbox{\usebox{\fmbox}}}

\raggedbottom
\onecolumn

\begin{center}
{\Large \bf
Family of nonstandard integrable and superintegrable classical Hamiltonian systems in non-vanishing magnetic fields\\
}
\vspace{6mm}
{Md Fazlul Hoque$^{a,b}$  and  Libor \v{S}nobl$^{a}$
}
\\[6mm]
\noindent ${}^{a}${\em 
Czech Technical University in Prague, Faculty of Nuclear Sciences and Physical Engineering,
 Department of Physics, B\v{r}ehov\'a 7, 115 19 Prague 1, Czech Republic}
\\[3mm]
\noindent ${}^{b}${\em
Pabna University of Science and Technology, Faculty of Science, Department of Mathematics, Pabna 6600, Bangladesh}\\ [1mm]
\vspace{4mm}
{\footnotesize Email: fazlulmath@pust.ac.bd and libor.snobl@fjfi.cvut.cz}
\end{center}

\vskip 1cm
\begin{abstract}
\noindent In this paper, we present the construction of all nonstandard integrable systems in magnetic fields whose integrals have leading order structure corresponding to the case (i) of Theorem 1 in [A Marchesiello and L \v{S}nobl 2022 {\it J. Phys. A: Math. Theor.} {\bf 55} 145203]. We find that the resulting systems can be written as one family with several parameters. For certain limits of these parameters the system belongs to intersections with already known standard systems separating in Cartesian and / or cylindrical coordinates and the number of independent integrals of motion increases, thus the system becomes minimally superintegrable. These results generalize the particular example presented in section 3 of [A Marchesiello and L \v{S}nobl 2022 {\it J. Phys. A: Math. Theor.} {\bf 55} 145203].
\end{abstract}

\section{Introduction}
The purpose of this study is to search for {classical} integrable and superintegrable systems in non-vanishing magnetic fields on the three-dimensional Euclidean space based on the class (i) of Theorem 1 in \cite{Sno1}, that is, on a non-standard pair of commuting quadratic elements of the universal enveloping algebra of the three-dimensional Euclidean algebra. 

Let us recall that a time independent $n$-dimensional Hamiltonian system given by the Hamiltonian $H$ is Liouville integrable if it admits $n$ integrals of motion $X_i$, $i=1,\dots, n$ (including the Hamiltonian) that are in involution with $H$ and are mutually in involution with each other. The integrable system is superintegrable if it allows additional integrals of motion $Y_j$, $j=1,\dots,k\leq n-1$ that are in involution with $H$ and not necessarily in involution with all others. The system is minimally superintegrable if it has $n+1$ integrals and maximally superintegrable if it has {$2n-1$} integrals including the Hamiltonian. The Kepler system and the isotropic harmonic oscillators are the best known examples of maximally superintegrable systems. As superintegrable systems admit more integrals of motion they afford {deeper} quantitative insight into the dynamical systems under examination { and both in the classical and quantum context find diverse applications in modern physics and many domains of pure and applied mathematics \cite{mil13}. For example, it is conjectured that all quantum maximally superintegrable systems are exactly solvable~\cite{TemTurWin}, although the exact scope of its validity (e.g. possible restriction on the polynomial order of the integrals) was so far not elucidated.}

Modern theory of superintegrability was first introduced in {the} mid--sixties by Winternitz et. al. \cite{Win1, Win2} and a systematic classification of {classical and quantum} two-- and three--dimensional superintegrable systems involving quadratic integrals of motion was completely studied and solved on conformally flat spaces \cite{Evan1, Kal2005, Kal2006, Kal2006a}. However, these studies were restricted to systems with only scalar potentials. {Recently, more general settings attracted substantial interest. These involved among others: study of quantum superintegrability involving spins, e.g. involving spin--orbital coupling~\cite{Win3} or possessing generalized Laplace--Runge--Lenz vectors~\cite{Nik1};
 superintegrability of classical and quantum systems in the magnetic monopole field with various scalar potentials \cite{Wu1, Jac1, Lab1, FH1, FH2, FH3}; and 
 the general search for superintegrable systems in non--vanishing magnetic fields which was extensively investigated from various directions on two--dimensional spaces \cite{Win4,Ber1, Cha1, Puc1, Puc2}.
}

Recently, {integrable and} superintegrable systems in the magnetic field in three dimensional Euclidean space were constructed by applying certain assumptions on the general structure of first and second-order integrals of motion corresponding to the leading order structures present in the absence of the magnetic field and related to {the} separation of coordinates in an associated orthogonal coordinate system\cite{Mar1}. Additional integrals of motion at most quadratic in momenta were also investigated for these systems to make {them} maximally or minimally superintegrable.  Four distinct classes of the second order spherical type classical integrable systems in a magnetic field have been established involving magnetic monopole as a special case \cite{Mar3}. Infinite families of cylindrical type integrable systems depending on arbitrary functions or parameters were presented in \cite{Fou1}. Also, three dimensional quadratically integrable systems in magnetic fields which possess non-subgroup type quadratic integrals have been studied \cite{Br1}. Superintegrability {and separability} of the systems in magnetic field {with} the integrals at most quadratic in the momenta which admit at least one cyclic coordinate, were investigated in \cite{Kub2021}. 

However, while investigating systems with Cartesian type integrals \cite{Mar2, Mar4, Mar5} it was observed that in the presence of magnetic fields, the leading order structure of the integrals may not always reduce {to} the classes relevant in the context of purely scalar potentials. This observation was further elaborated on in \cite{Sno1} where the classification of the leading order terms as admitted by the algebraic structure of the Euclidean algebra was performed and in \cite{Kub2022} where the systems with generalized cylindrical and spherical type integrals were classified, leading to several new classes of quadratically integrable systems with magnetic fields. In the present paper we focus on {one} class of commuting elements in the universal enveloping algebra of the Euclidean algebra, namely class (i) of \cite{Sno1} and classify all systems with the corresponding integrals. In addition to already known superintegrable systems (which are {at} the intersection with other, already investigated ``standard'' classes), {we find} one new class of nonseparable quadratically integrable systems.

The paper is organized as follows: Section \ref{sec2} contains {the} general structure of integrable Hamiltonian systems in magnetic fields with integrals of motion at most quadratic in the momenta and the assumed structure of integrals we shall focus on in the present paper {once the necessary notation is introduced. The main result of~\cite{Sno1} is recalled and also the results of the present paper are summarized as a theorem.} In section \ref{sec3}, we express the conditions on the Hamiltonian systems and {their} integrals of motion in the cylindrical coordinates. Sections \ref{sec4} and \ref{sec5} {present in some detail the derivation of the resulting} families of nonstandard integrable and superintegrable systems in the non--vanishing magnetic fields {together with explicit structure of their integrals of motion}. Finally, section \ref{sec6} presents the conclusion and further discussion.

\section{The three-dimensional classical Hamiltonian systems and integrals of motion in magnetic fields} \label{sec2}

Let us consider a charged particle moving in a static electromagnetic field in the three-dimensional Euclidean space which is described by a classical Hamiltonian system,
\begin{eqnarray}
H(\vec{x},\vec{p})=\frac{1}{2}\left(\vec{p}+\vec{A}(\vec{x})\right)^2+W(\vec{x}),\label{chm1}
\end{eqnarray}
where $\vec{p}=(p_1,p_2,p_3)$ are components of the linear momentum and $\vec{x}=(x_1,x_2,x_3)\equiv (x,y,z)$ the Cartesian spatial coordinates, $\vec{A}(\vec{x}) =(A_1(\vec{x}), A_2(\vec{x}), A_3(\vec{x})) \equiv (A_x(\vec{x}), A_y(\vec{x}), A_z(\vec{x}))$ is the vector potential depending on the position vector $\vec{x}$ and  $W(\vec{x})$ is the electrostatic potential function involving only the coordinates $\vec{x}$. We choose the units in which the mass of the particle has the numerical value 1 and the charge of the particle is $-1$.
The dynamics is invariant under time--independent gauge transformation
\begin{eqnarray}
\vec{A}'(\vec{x})=\vec{A}(\vec{x})+\nabla\chi, \qquad W'(\vec{x})=W(\vec{x}).\label{aw1}
\end{eqnarray}
The physically relevant magnetic field $\vec{B}(\vec{x})$ can be derived from the curl of the vector potential $\vec{A}(\vec{x})$, i.e., the relation 
\begin{eqnarray}
\vec{B}(\vec{x})=\nabla \times \vec{A}(\vec{x}),\quad \text{i.e.}\quad B_j(\vec{x})={\sum_{k,l=1}^{3}} \epsilon_{jkl}\frac{\partial A_l}{\partial x_k}, \quad {j=1,2,3},\label{mgc1}
\end{eqnarray}
where $\epsilon_{jkl}$ is the completely antisymmetric tensor with $\epsilon_{123}=1$. The scalar potential of the Hamiltonian system (\ref{chm1}) reads
\begin{eqnarray}
V(\vec{x})=W(\vec{x})+\frac{1}{2}\left(A_1(\vec{x})^2+A_2(\vec{x})^2+A_{3}(\vec{x})^2\right),
\end{eqnarray}
which is affected by the gauge transformation relation (\ref{aw1}). We will search for integrable and superintegrable Hamiltonian systems in non-vanishing magnetic fields, i.e., at least one of the components of the magnetic field is assumed to be nonzero.

A three-dimensional classical Hamiltonian system is said to be (Liouville) integrable, if it allows two integrals of motion {$X_1$ and $X_2$}, which are in involution with the Hamiltonian $H$, that is,
\begin{eqnarray}
{\{X_1,H\}_{P.B.}=0=\{X_2,H\}_{P.B} },
\end{eqnarray}
where the Poisson bracket $\{\quad,\quad\}_{P.B}$ is defined {on the phase space} by
\begin{eqnarray}
 \{\mathcal{X}(\vec{x},\vec{p}),\mathcal{Y}(\vec{x},\vec{p})\}_{P.B}=\sum_{k=1}^3\left(\frac{\partial\mathcal{X}}{\partial x_k}\frac{\partial\mathcal{Y}}{\partial p_k}-\frac{\partial\mathcal{Y}}{\partial x_k}\frac{\partial\mathcal{X}}{\partial p_k} \right).
\end{eqnarray}
It is also required that they are mutually in involution, that is, {$\{X_1,X_2\}_{P.B.}=0$}. Moreover, the Hamiltonian $H$ and the integrals of motion {$X_1$, $X_2$} must be functionally independent. That means that the rank of the matrix 
\begin{eqnarray}
\left[\frac{\partial (H,X_1,X_2)}{\partial (x_k,p_k)}\right]
\end{eqnarray}
must be 3 for the three--dimensional case. A classical integrable Hamiltonian system is superintegrable if it admits at least one additional functionally independent integral of motion. A three-dimensional superintegrable system is minimally superintegrable if it has exactly four functionally independent integrals and it is maximally superintegrable if it admits five functionally independent integrals of motion including the Hamiltonian.

{Classifying integrable systems one often assumes that the integrals are polynomial in the momenta, typically at most quadratic. The condition that we have integrals of motion in involution, i.e.,
\begin{eqnarray}
\{H,X_1\}_{P.B.}=\{H,X_2\}_{P.B.}= \{X_1,X_2\}_{P.B.}=0 \label{pcr1}
\end{eqnarray}
implies that the leading order terms in the integrals are commuting quadratic  expressions constructed out of linear and angular momenta $p_i$ and $l_i$, i.e. commuting quadratic elements in the universal enveloping algebra $\mathfrak{U}(\mathfrak{e}_3)$ {of the Euclidean algebra}. Thus we can use as our starting point the main result of~\cite{Sno1}, namely the theorem stating the following:
\begin{theorem}\label{EnvAlg_MainTheor}
Any three--dimensional Abelian subalgebra $\mathrm{span}\{h,X_1,X_2\}$ of quadratic commuting elements in the universal enveloping algebra $\mathfrak{U}(\mathfrak{e}_3)$ can be modulo the equation 
$$\vec p \cdot \vec l=\sum_{j=1}^3 p_j l_j=0$$ 
and transformations from the Euclidean group written in terms of the following elements
\begin{enumerate}[(a)]
\item \label{speroblprol}\begin{equation*}
X_1=l_1^2+l_2^2+l_3^2+a l_3 p_3 + b p_3^2,\quad X_2=l_3^2, \quad a,b\in\mathbb{R},
\end{equation*}
\item \label{conicellipsoidal} \begin{equation*}
 X_1 = l_1^2+l_2^2+l_3^2+b (a p_2^2+ p_3^2 ), \quad X_2 = a l_2^2+l_3^2-a b p_1^2, \quad 0< a \leq \frac{1}{2},b\in\mathbb{R},
\end{equation*}
\item \label{strangeI} \begin{eqnarray*}
\nonumber  X_1 & = & l_1^2+l_2^2+l_3^2+ 2 b ( l_1 p_1- (3 a-1) l_2 p_2-2 l_3 p_3)+\\
& & + 3  b^2 ( (1-4 a) p_1^2 - (3 a^2-2 a-1) p_2^2+ 2 (a-1) p_3^2),\\ 
\nonumber   X_2 & = & a l_2^2+l_3^2+6 a b l_1 p_1+9 a b^2 (a p_3^2+p_2^2) , \quad 0< a \leq \frac{1}{2}, b\in\mathbb{R} \backslash \{0 \},
\end{eqnarray*}
\item\label{circularparabolic} \begin{equation*}
 X_1=l_3^2,\quad X_2=\frac{1}{2}\left( l_1 p_2+ p_2 l_1 - l_2 p_1-p_1 l_2\right)+a l_3 p_3,\;  a\geq 0,
\end{equation*}
\item\label{paraboloidal}
\begin{equation*}
  X_1=l_3^2+2 a (l_1 p_1-l_2 p_2) + a^2 p_3^2, \quad
X_2=\frac{1}{2} \left(l_1 p_2+p_2 l_1 -l_2 p_1-p_1 l_2 \right)-a p_1 p_2, \;   a >0,
\end{equation*}
\item\label{cylindrical}
\begin{equation*}
 X_1=l_3^2+a l_3 p_3+b p_1^2 +c p_1 p_3+d p_2 p_3, \quad X_2=p_3^2, \quad a,b\in\mathbb{R}, \,  c\geq 0, \,  d\geq 0,
\end{equation*}
\item\label{strangeII}
\begin{equation*}
 X_1=l_3^2+a p_3^2 ,\quad X_2= l_3 p_3+b p_3^2, \quad a,b\in\mathbb{R},
\end{equation*}
\item\label{strangeIV}\begin{equation*}
 X_1 = l_1 p_1+a l_2 p_2-(a+1) l_3 p_3+b p_2^2, \quad 
X_2 = p_1^2+\frac{2 a+1}{a+2} p_2^2, \quad -\frac{1}{2}< a \leq 0, b\in\R,
\end{equation*}
\item\label{strangeV}\begin{equation*}
 X_1 = l_1 p_1+a p_2^2+b p_2 p_3, \quad X_2 = p_1^2, \quad a\in\R,\,  b\geq 0,
\end{equation*}
\item\label{ParabolicCylindrical}
\begin{eqnarray*}
\nonumber X_1 & = & l_1 p_1+a l_2 p_2-(a+1) l_3 p_3+\frac{\omega}{2} \left(l_1 p_3+p_3 l_1 -l_3 p_1-p_1 l_3 \right)+2 b  p_1 p_2+c \left(p_2^2- p_3^2\right),\\
 X_2 & = & p_1^2+ \frac{6 \omega}{4 a-1}  p_1 p_3+\frac{a+2}{4 a-1} p_2^2-\frac{5 a+1}{4 a-1} p_3^2,
\end{eqnarray*} 
where $\omega = \sqrt{1+a-2 a^2}$, $-\frac{1}{2}< a\leq 0$, $b \geq 0$, $c\in\mathbb{R}$,
\item\label{Cartesian}\begin{equation*}
 X_1=p_1^2+a p_2^2, \quad X_2=p_2^2 + b p_1 p_2+c p_1 p_3+ d p_2 p_3, \quad 0\leq a \leq \frac{1}{2}, \, b\geq 0, \, c\geq 0, \,d\in\mathbb{R}.
\end{equation*} 
\end{enumerate}
\end{theorem}
Out of these classes, most occur in the classification of quadratically integrable systems with scalar potentials for some particular values of the parameters, cf.~\cite{Win2}. These systems also coincide with the systems allowing orthogonal separation of variables in the Hamilton--Jacobi equation, cf.~\cite{EisenhartAnnMath}. However, the classes (c) and (g)--(i) were not encountered before and their relevance for the construction of integrable systems is not yet clear. 

In order to elucidate their relevance, we shall focus here on the class (i) of the Theorem~\ref{EnvAlg_MainTheor}. After a rotation respecting the physicists' preference for alignment along the $z$--axis, the leading order structure of the integrals of motion of the Hamiltonian system as in Theorem~\ref{EnvAlg_MainTheor}
implies the general form of our quadratic integrals of motion as}
\begin{eqnarray}
&&X_1= l^A_3p_3^A+a(p^A_1)^2+bp^A_1p_2^A+\sum_{j=1}^3 s_{j}(\vec{x}) p_j^A+m(\vec{x}),\label{int1}
\\&&
X_2=(p^A_3)^2+\sum_{j=1}^3 S_{j}(\vec{x})p^A_j+M(\vec{x}),\quad a,b\in \mathbb{R}, \label{int2}
\end{eqnarray}
where the gauge covariant expressions were used {for convenience}\footnote{{Use of the covariant expressions $p_i^A$ and $l_i^A$ instead of $p_i$ and $l_i$  amounts to a suitable redefinition of the lower order terms and does not imply any loss of generality.}
}, namely
\begin{eqnarray}
p_i^A=p_i+A_i(\vec{x}),\quad l_i^A=\sum_{j,k=1}^{3}\varepsilon_{ijk}x_jp^A_k, \quad {i=1,2,3}.
\end{eqnarray}
Here $\varepsilon_{ijk}$ is the completely antisymmetric tensor with $\varepsilon_{123}=1$. {The functions $s_j(\vec x)$, $S_j(\vec x)$, $m(\vec x)$ and $M(\vec x)$ are so far arbitrary and need to be determined from the condition that the integrals of motion $X_i$, $i=1,2$ {are} in involution with the Hamiltonian $H$} and mutually, i.e., {equations~\eqref{pcr1}, considered as polynomials in the momenta; thus implying that the coefficient of each monomial in the momenta must vanish.}

{In order to avoid {the} use of multiple indices, we shall use the notation that lower--case functions like  $s_j(\vec x)$, integration constants etc. are related to the first integral $X_1$, the upper--case ones like $S_j(\vec x)$ refer to the second integral $X_2$ and greek letters are used for a hypothetical additional integral(s), see below.}

For the integrable systems found, we search for additional at most quadratic integrals of motion of the {most general form allowed by the leading order terms in the condition $\{H,X_3\}_{P.B.}=0$ which reads}
\begin{eqnarray}
X_3=\sum_{1\leq i\leq j\leq 6} {\alpha_{ij}} T_i^A T_j^A + \sum_{k=1}^3 {\sigma_k}(\vec{x}) p^A_k + {\mu}(\vec{x}),\label{adsint}
\end{eqnarray}
where 
\begin{eqnarray}
T^A=(p_1^A,p_2^A,p_3^A,l_1^A,l_2^A,l_3^A), \quad a_{ij}\in \mathbb{R}.
\end{eqnarray}
The Poisson bracket of the integral of motion $X_3$ and the Hamiltonian $H$ must vanish, that is, we have the {remaining conditions implied by} $\{H,X_3\}_{P.B.}=0$ to determine the constants {$\alpha_{ij}$} and the a priori arbitrary functions {$\sigma_k(\vec{x})$} and {$\mu(\vec{x})$} {together with restrictions on the parameters in the magnetic field $\vec B(\vec x)$ and the electrostatic potential $W(\vec x)$.}

{Our main result reads

\begin{theorem}\label{NewTheorem}
There exists one and only one family of quadratically integrable Hamiltonian systems with the integrals $X_1$ and $X_2$ of the form~\eqref{int1}--\eqref{int2} which for generic values of its parameters does not possess any other independent quadratic integrals of motion and thus belongs to the class of systems characterized by the leading order structure (i) of Theorem~\ref{EnvAlg_MainTheor} and no other. In addition to the parameters $a,b$ characterizing the class (i) of Theorem~\ref{EnvAlg_MainTheor} it involves two arbitrary parameters $b_\phi$ and $b_Z$ in the magnetic field 
\begin{equation*}
\vec{B}(\vec{x})=(- b_\phi y, b_\phi x,b_Z),\nonumber
\end{equation*}
and three parameters $w_1$, $w_2$ and $w_3$ in the electrostatic potential
\begin{equation*}
 W(\vec{x})  = b_\phi \left(  -\frac{1}{8}   b_\phi(x^2+y^2)^2   +  w_3 (x^2+y^2) -{b_Z} \left( \frac{a}{2} \left( x^2 - y^2 \right)+ b x y \right) +  w_1 x + w_2 y \right).
\end{equation*}

In the limit of vanishing $b_\phi$ or $b_Z$ this system becomes quadratically superintegrable and obtains other commuting pairs of quadratic integrals, thus in these limits belongs to an intersection of several classes of Theorem~\ref{EnvAlg_MainTheor}.
\end{theorem}
}

In the following sections, we will {present in some detail our derivation of Theorem~\ref{NewTheorem}, write down explicit formulas for the integrals and also the Poisson algebra of integrals in the superintegrable limits}. For the sake of {computational} convenience, we consider our equations in {the} cylindrical coordinates.

\section{The Hamiltonian systems and integrals of motion in cylindrical coordinates}\label{sec3}
Let us now express the integrals of motion (\ref{int1})-(\ref{int2}) in the cylindrical coordinates. We introduce the cylindrical coordinates in a standard way,
\begin{eqnarray}
x=r\cos(\phi),\qquad y=r\sin(\phi), \qquad z=Z.
\end{eqnarray}
The linear momenta transform,
\begin{eqnarray}
p_1=\cos(\phi) p_r - \frac{\sin(\phi)}{r}p_\phi,\quad
 p_2=\sin(\phi) p_r + \frac{\cos(\phi)}{r}p_\phi,\quad
 p_3=p_Z \label{picy1}
\end{eqnarray}
and similarly the components of the vector potential 1-form
\begin{eqnarray}
A =A_1(\vec{x})dx_1+A_2(\vec{x})dx_2+A_3(\vec{x})dx_3=A_r dr+A_\phi  d\phi +A_Z dZ,
\end{eqnarray}
namely,
\begin{eqnarray}
&&A_1(\vec{x})=\cos(\phi)A_r(r,\phi, Z)-\frac{\sin(\phi)}{r}A_\phi (r,\phi, Z),\nonumber\\&&
A_2(\vec{x})=\sin(\phi)A_r(r,\phi, Z)+\frac{\cos(\phi)}{r}A_\phi (r,\phi, Z),\\&&
A_3(\vec{x})=A_Z(r,\phi, Z).\nonumber
\end{eqnarray}
Thus also the covariant momenta transform as
\begin{eqnarray}
 p_1^A= \cos(\phi)p^A_r-\frac{\sin(\phi)}{r}p^A_\phi,\quad
p_2^A= \sin(\phi)p^A_r+\frac{\cos(\phi)}{r}p^A_\phi,
\quad
p_3^A= p_Z^A.\label{pa1}
\end{eqnarray}
The 2-form of the magnetic field is related to the vector potential through its exterior derivative, $B=d A$, thus its components transform as
\begin{eqnarray}
&& B^x(\vec{x})=\frac{\cos(\phi)}{r} B^r(r,\phi, Z)-\sin(\phi)B^\phi (r,\phi, Z),\nonumber\\&&
B^y(\vec{x})=\frac{\sin(\phi)}{r}B^r(r,\phi, Z)+\cos(\phi)B^\phi (r,\phi, Z),\label{mag1}
\\&&
B^z(\vec{x})=\frac{1}{r} B^Z(r,\phi, Z).\nonumber
\end{eqnarray}
The Hamiltonian in the cylindrical coordinates reads
\begin{eqnarray}
H = \frac{1}{2}\left((p_r^A)^2+\frac{(p_\phi^A)^2}{r^2}+(p_Z^A)^2\right)+W(r,\phi, Z),\label{hmcy1}
\end{eqnarray}
where $p^A_\lambda=p_\lambda +A_\lambda(r,\phi, Z)$, $\lambda=r,\phi, Z$ and the integrals (\ref{int1})-(\ref{int2}) are written as
\begin{eqnarray}
&& X_1=p^A_\phi p^A_Z +a \left(\cos(\phi) p^A_r - \frac{\sin(\phi)}{r}p^A_\phi)\right)^2
+b\left(\cos(\phi) p^A_r - \frac{\sin(\phi)}{r}p^A_\phi \right)\nonumber\\&& \qquad \quad \times\left(\sin(\phi) p^A_r + \frac{\cos(\phi)}{r}p^A_\phi \right) +\sum_{\lambda= r,\phi,Z} s_{\lambda}(r,\phi,Z)p^A_\lambda + m(r,\phi,Z),\label{Y1}
\\&&
X_2= (p^A_Z)^2+\sum_{\lambda= r,\phi,Z} S_{\lambda}(r,\phi,Z)p^A_\lambda + M(r,\phi,Z).\label{Y2}
 \end{eqnarray}
The general structure of the second order integral (\ref{adsint}) reads
 \begin{eqnarray}
X_3=\sum_{1\leq i\leq j\leq 6} {\alpha}_{ij} T_i^A T_j^A + \sum_{\lambda=r,\phi, Z} {\sigma}_\lambda(r,\phi,Z) p^A_\lambda + {\mu}(r,\phi,Z),\label{X4}
 \end{eqnarray}
where in $T^A$ the linear momentum $p_i^A$ is given by (\ref{pa1}) and the covariant angular momentum is expressed as
\begin{eqnarray}
&&
l^A_1=- Z\sin(\phi)p^A_r -\frac{Z\cos(\phi)}{r} p^A_\phi + r\sin(\phi) p^A_Z,\nonumber
\\&&
l^A_2 = Z\cos(\phi) p^A_r - \frac{Z\sin(\phi)}{r}p^A_\phi - r\cos(\phi) p^A_Z, \\&&
 l^A_3=p^A_\phi.\nonumber
\end{eqnarray}

\subsection{The conditions for the integrals of motion}
In order to obtain all systems with the integrals of the form (\ref{Y1})-(\ref{Y2}), we need to solve the involutivity conditions
\begin{eqnarray}
\{H,X_1\}_{P.B.}=\{H,X_2\}_{P.B.}= \{X_1,X_2\}_{P.B.}=0,\label{pbr1}
\end{eqnarray}
where the Poisson bracket in the cylindrical coordinates reads 
\begin{eqnarray}
 \{{\mathcal{X}},{\mathcal{Y}}\}_{P.B}=\sum_{\lambda=r,\phi,Z}\left(\frac{\partial {\mathcal{X}}}{\partial \lambda}\frac{\partial {\mathcal{Y}}}{\partial p_\lambda}-\frac{\partial {\mathcal{Y}}}{\partial \lambda}\frac{\partial {\mathcal{X}}}{\partial p_\lambda} \right).
\end{eqnarray}
The conditions (\ref{pbr1}) can be viewed as polynomials in the momenta $p_\lambda$. Since they must hold for all values of the momenta, they {imply} the following equations as coefficients of the second, first and zeroth order  monomials in the momenta.
\begin{itemize}
\item[]
$\{H,X_1\}_{P.B.}=0:$\\
Second order:
\begin{eqnarray}
&&\partial_r s_{r}=-\frac{1}{r}\left(2b\cos^2(\phi)-2a\cos(\phi)\sin(\phi)-b\right)B_Z,\nonumber\\&&
\partial_r s_\phi=\frac{1}{r^2}\left(4a\cos^2(\phi) B_Z+4b\cos(\phi)\sin(\phi) B_Z+r^2B_\phi-2a B_Z-\partial_\phi s_r\right),\nonumber\\&&
\partial_\phi s_\phi=\frac{1}{r}\left(2b\cos^2(\phi) B_Z-2a\cos(\phi)\sin(\phi) B_Z-b B_Z-r B_r-s_r\right),\label{hy1s}
\\&&
\partial_\phi s_Z=-2br\cos^2(\phi) B_\phi-2a\cos^2(\phi) B_r +2ar\cos(\phi)\sin(\phi) B_\phi-2b\cos(\phi)\sin(\phi) B_r \nonumber\\&&\qquad\quad -r^2\partial_Z s_\phi+brB_\phi+2aB_r,\nonumber\\&&
\partial_Z s_r=\frac{1}{r}\left(2b\cos^2(\phi) B_r-2ar\cos^2(\phi) B_\phi-2br\cos(\phi)\sin(\phi) B_\phi -2a\cos(\phi)\sin(\phi) B_r\right. \nonumber\\&&\qquad\quad \left.- bB_r - rB_Z - r\partial_r s_Z\right),\nonumber\\&&
\partial_Z s_Z=B_r.\nonumber
\end{eqnarray}
\item[]
First order:
\begin{eqnarray}
&&\partial_Z m= s_\phi B_r - s_r B_\phi+\partial_\phi W,\nonumber\\&&
\partial_\phi m = 2br\cos^2(\phi)\partial_r W -2a\cos^2(\phi)\partial_\phi W-2ar\cos(\phi)\sin(\phi)\partial_r W \label{hy1m}
\\&&\qquad\quad -2b\cos(\phi)\sin(\phi)\partial_\phi W  +r^2\partial_Z W-br\partial_r W+s_r B_Z-s_Z B_r+2a\partial_\phi W,\nonumber
\\&&
\partial_r m=\frac{1}{r}\left(2ar\cos^2(\phi)\partial_r W+2b\cos^2(\phi)\partial_\phi W +2br\cos(\phi)\sin(\phi)\partial_r W\right. \nonumber\\&&\qquad\quad \left.-2a\cos(\phi)\sin(\phi)\partial_\phi W  -r s_\phi B_Z+r s_Z B_\phi - b \partial_\phi W\right).\nonumber
\end{eqnarray}
\item[]
Zeroth order:
\begin{eqnarray}
s_{r}\partial_r W+s_{\phi} \partial_\phi+s_{Z}\partial_Z W=0.\label{hy1w}
\end{eqnarray}
\end{itemize}
\begin{itemize}
\item[]
$\{H,X_2\}_{P.B.}=0:$\\
Second order:
\begin{eqnarray}
&&\partial_r S_r=0, \quad \partial_r S_\phi=-\frac{1}{r^2}\partial_\phi S_r,\quad  \partial_\phi S_\phi=-\frac{1}{r}S_r,\nonumber\\&& \partial_\phi S_Z=-r^2\partial_Z S_\phi-2B_r, \quad \partial_Z S_Z=0,\quad \partial_Z S_r=2B_\phi-\partial_r S_Z. \label{hy1S}
\end{eqnarray}
\item[]
First order:
\begin{eqnarray}
&&\partial_r M=S_Z B_\phi-S_\phi B_Z, \quad \partial_\phi M=S_r B_Z-S_Z B_r, \label{hy2m}
\\&& \partial_Z M=-S_r B_\phi+S_\phi B_r+2\partial_Z W.\nonumber
\end{eqnarray}
\item[]
Zeroth order:
\begin{eqnarray}
S_{r}\partial_r W+ S_{\phi}\partial_\phi W+S_{Z}\partial_Z W=0.\label{hy2w}
\end{eqnarray}
\end{itemize}
$\{X_1,X_2\}_{P.B.}=0:$\\
Second order:
\begin{eqnarray}
&&\Big(8ab\cos^4(\phi) -4 (a^2 -b^2)\cos^3(\phi)\sin(\phi) - 6ab\cos^2(\phi) - 2b^2\cos(\phi)\sin(\phi)\Big)\partial_{\phi} S_r \nonumber\\&& +(2b\cos^2(\phi) - 2 a\cos(\phi)\sin(\phi) - b )r^2\partial_{r}S_Z -\Big(8ab\cos^4(\phi) - 4(a^2 - b^2)\cos^3(\phi)\sin(\phi) \nonumber\\&& - 6ab\cos^2(\phi) - 2b^2\cos(\phi)\sin(\phi)\Big)r S_\phi -\Big(4b\cos^2(\phi) - 4a\cos(\phi)\sin(\phi) \nonumber\\&&- 2b\Big)r^2B_\phi=0,\nonumber\\&&
\Big(8(b^2-3a^2)br\cos^6(\phi) +8(a^2 - 3b^2)ar\cos^5(\phi)\sin(\phi) + 4(5a^2 - 3b^2)br\cos^4(\phi) \nonumber\\&&+ 16 r a b^2\cos^3(\phi)\sin(\phi) + 4\cos^2(\phi)r b^3\Big)\partial_\phi S_r-\Big( 8 a br^3\cos^4(\phi)- 4(a^2 \nonumber\\&& -b^2)r^3 \cos^3(\phi)\sin(\phi)  - 6 a b r^3\cos^2(\phi)- 2 b^2 r^3\cos(\phi)\sin(\phi)\Big)\partial_r S_Z \nonumber\\&& +b^2\Big( 4b\cos^2(\phi) - 4a\cos(\phi)\sin(\phi) - 2b \Big)r\partial_r s_Z + 4 b^2(b\sin(\phi) + a\cos(\phi))\cos(\phi)r^2\partial_Z s_\phi \nonumber\\&& +\Big(8(3 a^2 - b^2)b\cos^6(\phi) -8 (a^2 - 3b^2) a\cos^5(\phi)\sin(\phi) - 4(5a^2 - 3b^2)b\cos^4(\phi) \nonumber\\&& - 16 a b^2\cos^3(\phi)\sin(\phi) - 2 b^3\cos^2(\phi) - 2ab^2 \cos(\phi)\sin(\phi) -  b^3 \Big)r^2S_\phi\nonumber\\&& +\Big( 16(b^2 +r^2)abr\cos^4(\phi) - 8( a^2 b^2 +  a^2 r^2 - b^4 -  b^2 r^2)r\cos^3(\phi)\sin(\phi)\nonumber\\&& -12 ( b^2 + r^2)abr\cos^2(\phi) -4(b^2 + r^2)b^2r\cos(\phi)\sin(\phi) \Big)B_\phi +8b^2\Big((a^2 - b^2)\cos^4(\phi)\nonumber\\&& + 2ab\cos^3(\phi)\sin(\phi) -(a^2 -b^2)\cos^2(\phi) - a b\sin(\phi)\cos(\phi) \Big)B_r +2b^2\Big(2b\cos^2(\phi) \nonumber\\&&- 2a\cos(\phi)\sin(\phi) - b\Big)r B_Z+2b^2(a\cos^2(\phi)+ b\cos(\phi)\sin(\phi))rS_r=0,\label{y1y2s}
\end{eqnarray}
\begin{eqnarray}
&&r^2\partial_Z S_\phi +2B_r=0,
\nonumber\\&&
\Big(4(a^2 - b^2)\cos^4(\phi) + 8ab\cos^3(\phi)\sin(\phi) + 4b^2\cos^2(\phi)\Big)\partial_\phi S_r+ 2(a\cos(\phi)\nonumber\\&& +b\sin(\phi))r^2\cos(\phi)\partial_r S_Z-2b^2\partial_r s_Z+\Big(4(b^2-a^2)\cos^4(\phi) - 8  a b\cos^3(\phi)\sin(\phi)\nonumber\\&& - 4 b^2\cos^2(\phi) - b^2\Big)rS_\phi -4(b \sin(\phi)+a\cos(\phi))r^2\cos(\phi)B_\phi  -2b^2 B_Z \nonumber\\&& -2b^2(b\sin(\phi) + a\cos(\phi))\cos(\phi)\partial_r S_Z=0,
\nonumber\\&&
(2b\cos^2(\phi) -2a\cos(\phi)\sin(\phi))\partial_\phi S_r -2br\cos^2(\phi)S_\phi +2a\cos(\phi)\sin(\phi)rS_\phi -b\partial_\phi S_r\nonumber\\&& +brS_\phi-r^2\partial_Z S_\phi=0,
\nonumber\\&&
 -16\Big((a^2 - b^2)\cos^2(\phi) + 2ab\sin(\phi)\cos(\phi) + b^2\Big)\cos^2(\phi)\partial_\phi S_r -8\Big( a\cos^2(\phi) \nonumber\\&& +b\cos(\phi)\sin(\phi)\Big)r^2\partial_r S_Z +\Big(16(a^2 - b^2)\cos^4(\phi) + 32ab\cos^3(\phi)\sin(\phi)\nonumber\\&& + 16b^2\cos^2(\phi) + 4 b^2 \Big)rS_\phi + 16(a\cos^2(\phi)+ b\cos(\phi)\sin(\phi))r^2B_{\phi}-4b^2\partial_\phi S_r=0.\nonumber
\end{eqnarray}
The lower order equations for $\{X_1,X_2\}_{P.B.}=0$ can be derived in the same way but are a bit too cumbersome to display here.

We have to split our computation into two cases:
\begin{itemize}
\item[$\bullet$] at least one of the parameters $a, b$ is nonvanishing, i.e., $a^2+b^2\neq 0$,
\item[$\bullet$] $a=b=0$, 
\end{itemize}
for which the equations (\ref{hy1s})-(\ref{y1y2s}) are structurally somewhat different (namely, for $a=b=0$ some equations vanish identically).
\section{The systems and integrals of motion for $a^2+b^2\neq 0$}\label{sec4}

\subsection{The general structure of $s_\lambda(r,\phi,Z)$, $S_\lambda(r,\phi,Z)$ and $B_\lambda(r,\phi,Z)$}

The equations (\ref{hy1s}), (\ref{hy1S}) and (\ref{y1y2s}) as equations for $s_\lambda(r,\phi,Z)$, $S_\lambda(r,\phi,Z)$ and $B_\lambda(r,\phi,Z)$ can be solved in full generality and imply the following structure
\begin{eqnarray}
&& 
s_r(r,\phi,Z)=-\frac{16}{15} \Big(9 S_{Z2} r^5 +5 S_{Z3}  r^3\Big)a b\cos^4(\phi) +\frac{8}{15}(a^2-b^2)(9 S_{Z2} r^2\nonumber\\&&\qquad + 5S_{Z3})\cos^3(\phi)\sin(\phi) +\frac{1}{15}\Big(144S_{Z2} a b r^5  + 10 b(10 S_{Z3}a + s_{Z2})r^3 + 30b(S_{Z4}a \nonumber\\&&\qquad + s_{Z1})r\Big)\cos^2(\phi) -\frac{1}{15}\Big(36S_{Z2}(a^2 -b^2)r^5 + 10(4S_{Z3}a^2 - 2S_{Z3}b^2 +  s_{Z2}a)r^3\nonumber\\&&\qquad + 30(S_{Z4}a  + s_{Z1})a r\Big)\cos(\phi)\sin(\phi)+\frac{1}{2}(2 s_{r2}+S_{r1}Z)\cos(\phi) +\frac{1}{2} (2s_{r1}\nonumber\\&&\qquad -S_{r2}Z)\sin(\phi)  -\frac{6}{5}S_{Z2} a b r^5 -\frac{1}{3} (s_{Z2} +4 S_{Z3}a)br^3-(S_{Z4}a +s_{Z1})br, \label{hsS1}
\\
&& 
s_\phi(r,\phi,Z)=\frac{1}{15}\Big(108 S_{Z2}r^2+40S_{Z3}\Big)(a^2 - b^2)r^2\cos^4(\phi)+\frac{1}{15}\Big(216 S_{Z2} r^4\nonumber\\&&\qquad +80S_{Z3}  r^2\Big)ab\cos^3(\phi)\sin(\phi)+\frac{1}{15}\Big(108 S_{Z2}(b^2 - a^2)r^4  -20(s_{Z2}a + 4S_{Z3}a^2 \nonumber\\&&\qquad - 2S_{Z3}b^2)r^2 -30S_{Z4}a^2 - 30s_{Z1}a\Big)\cos^2(\phi)-\frac{1}{15}\Big( 108 S_{Z2}a r^4 +20(4S_{Z3}a  +s_{Z2})r^2\nonumber\\&&\qquad  + 30S_{Z4}a + 30s_{Z1}\Big)b\cos(\phi)\sin(\phi)  + \frac{1}{2r}( 2 s_{r1}  -S_{r2}Z )\cos(\phi)\nonumber\\&&\qquad -\frac{1}{2r} (2s_{r2}+S_{r1}Z)\sin(\phi) + \frac{1}{2}S_{Z2}r^6  +\frac{1}{10}\Big((24a^2 + 6b^2)S_{Z2} +5 S_{Z3}\Big)r^4\nonumber\\&&\qquad +\frac{1}{6}\Big((8a^2 + 2b^2)S_{Z3} +4 s_{Z2}a +3 S_{Z4} \Big)r^2   + (S_{Z4}a + s_{Z1})a + s_{\phi 1},\nonumber
\\&&
s_Z(r,\phi,Z)=-(3 S_{Z2} r^6 + 2S_{Z3}r^4 + S_{Z4}r^2)(a\cos(\phi)+b\sin(\phi))\cos(\phi) -\frac{1}{2} S_{r1}r\cos(\phi)\nonumber\\&&\qquad +\frac{1}{2} S_{r2}r\sin(\phi)+ S_{Z2}ar^6 +\frac{1}{4}(4S_{Z3}a + s_{Z2})r^4 + \frac{1}{2}(S_{Z4}a + s_{Z1})r^2 + s_{Z3},\nonumber
\\&&
S_r(r,\phi,Z)=S_{r1}\sin(\phi) + S_{r2}\cos(\phi),\nonumber
\\&&
S_\phi(r,\phi,Z)=\frac{1}{r}\Big(S_{r1}\cos(\phi) - S_{r2}\sin(\phi)\Big),\nonumber
\\&&
S_Z(r,\phi,Z)=S_{Z2}r^6 + S_{Z3}r^4 + S_{Z4}r^2 + S_{Z1},\nonumber
\end{eqnarray}
where $s_{r1}, s_{r2}, s_{\phi 1}$, $s_{Z1}, s_{Z2}, s_{Z3}$, $S_{r1}, S_{r2}$, $S_{Z1}, S_{Z2}, S_{Z3}, S_{Z4}$ are integration constants. Thus, in contrast to the construction of integrable systems with  ``standard'' type integrals, e.g., Cartesian \cite{Mar2,Mar5}, cylindrical \cite{Fou1} or spherical \cite{Mar3}, the functions $s_\lambda(r,\phi,Z)$ and $S_\lambda(r,\phi,Z)$ do not depend on any arbitrary functions, they involve only integration constants.
The components of the magnetic field are also explicitly given in terms of these constants and read
\begin{eqnarray}
&& B_r(r,\phi,Z)=0,\quad
 B_\phi(r,\phi,Z)=3S_{Z2}r^5 + 2S_{Z3}r^3 + S_{Z4}r,\label{mags1}
\\&&
 B_Z(r, \phi, Z) = 4(3S_{Z2}r^2 + S_{Z3})ar^3\cos^2(\phi) +4 (3S_{Z2}r^2 + S_{Z3})br^3\cos(\phi)\sin(\phi)\nonumber\\&&\qquad - 6S_{Z2}ar^5 - (4S_{Z3}a + s_{Z2})r^3 -(S_{Z4}a + s_{Z1})r.\nonumber
\end{eqnarray}

\subsection{The solution of lower order equations}
To obtain the functions $m(r,\phi, Z)$ and $M(r,\phi, Z)$ in the integrals of motion (\ref{Y1}) and (\ref{Y2}), we now substitute the general forms of the coefficients $s_\lambda(r,\phi,Z)$ and $S_\lambda(r,\phi,Z)$, $\lambda=r,\phi, Z$ (\ref{hsS1}) of the second-order integrals and the magnetic field (\ref{mags1}) into the remaining equations. Using the compatibility conditions for the existence of the functions $m(r,\phi,Z)$ and $M(r,\phi,Z)$ coming from (\ref{hy1m}) and (\ref{hy2m}), we arrive at a system of equations involving the electrostatic potential $W(r,\phi,Z)$ and all the constants introduced so far. Some of these equations form an overdetermined system of PDEs for $W(r,\phi,Z)$, some do not involve $W(r,\phi,Z)$ and impose algebraic constraints on our constants as coefficients of various independent monomials in $r$, $Z$, $\sin(\phi)$ and $\cos(\phi)$ (taking $\sin^2(\phi)+ \cos^2(\phi)=1$ into account). The assumption of {a} nonvanishing magnetic field imposed on the possible solutions of the algebraic constraints and the solvability conditions for the PDEs for the potential $W(r,\phi,Z)$ {imply} the vanishing values of the following parameters
\begin{eqnarray}
S_{Z2} = 0,\quad S_{Z3} = 0,\quad S_{r1} = 0,\quad S_{r2} = 0,\quad s_{Z2} = 0.\label{CS1}
\end{eqnarray}
Applying these conditions to the general structure of $s_\lambda(r,\phi,Z)$ and $S_\lambda(r,\phi,Z)$ (\ref{hsS1}), it reduces to
\begin{eqnarray}
&& s_r(r, \phi, Z) =2 (S_{Z4}a + s_{Z1})br\cos^2(\phi) -2(S_{Z4}a +s_{Z1})ar\cos(\phi)\sin(\phi)\nonumber\\&&\qquad\qquad + s_{r2}\cos(\phi) + s_{r1}\sin(\phi) -(S_{Z4}a + s_{Z1}) b r,\nonumber
 \\&&  
  s_\phi(r, \phi, Z) = -2a(S_{Z4}a +s_{Z1})\cos^2(\phi) -2b (S_{Z4}a  + s_{Z1})\cos(\phi)\sin(\phi)\nonumber\\&&\qquad\qquad +\frac{1}{r}( s_{r1}\cos(\phi) - s_{r2}\sin(\phi))+\frac{1}{2} S_{Z4}r^2 + (S_{Z4} a + s_{Z1})a + s_{\phi 1}   ,\label{src}
 \\&& 
 s_Z(r, \phi, Z) = - S_{Z4}ar^2\cos^2(\phi) -S_{Z4}br^2\sin(\phi)\cos(\phi) +\frac{1}{2} (S_{Z4}a + s_{Z1})r^2 + s_{Z3},  \nonumber 
\\&&
S_r(r, \phi, Z) = 0,\nonumber
\\&&
 S_\phi(r, \phi, Z) = 0,\label{Src}
 \\&&
  S_Z(r, \phi, Z) = S_{Z4}r^2 + S_{Z1}.\nonumber
 \end{eqnarray}
The magnetic field takes the simple {two--parameter} form
 \begin{eqnarray}
B_r(r, \phi, Z) = 0,\quad B_\phi(r, \phi, Z) = S_{Z4}r,\quad  B_Z(r, \phi, Z) = \tilde{s}_{Z1}r,\label{fmg1}
\end{eqnarray}
where $\tilde{s}_{Z1}= -(S_{Z4}a + s_{Z1})$. The electrostatic potential $W(r,\phi,Z)$ is uniquely determined by the values of the parameters in (\ref{src})-(\ref{Src}) and reads
\begin{eqnarray}
&&W(r, \phi, Z) = -\frac{1}{8} S_{Z4}^2r^4  +\frac{1}{2} S_{Z4}\left(\tilde{s}_{Z1}a-s_{\phi 1}\right)r^2 -S_{Z4}\tilde{s}_{Z1}ar^2\cos^2(\phi) \nonumber\\&&\qquad\qquad  - S_{Z4}\tilde{s}_{Z1}br^2\sin(\phi)\cos(\phi)  - S_{Z4}s_{r1}r\cos(\phi) + S_{Z4}s_{r2}r\sin(\phi)\label{fpw1}
\end{eqnarray}
(dropping {an} irrelevant additive constant). {Introducing a simpler parametrization, namely calling $b_\phi$ and $b_Z$ the parameters present in the magnetic field and $w_i$ present only in the electrostatic potential, i.e. $b_\phi=S_{Z4}$, $b_Z=\tilde{s}_{Z1}$, $w_1=-s_{r1}$, $w_2=s_{r2}$, $w_3=-\frac{1}{2} s_{\phi 1}$, the fields take the form
\begin{eqnarray}
B_r(r, \phi, Z) = 0,\quad B_\phi(r, \phi, Z) = b_{\phi}r,\quad  B_Z(r, \phi, Z) = b_Z r\label{conmg1}
\end{eqnarray}
and the electrostatic potential
\begin{align}
&&W(r, \phi, Z) = {b_{\phi} \left( -\frac{1}{8} b_{\phi}r^4  + w_3 r^2 - \frac{b_Z }{2} \left( a \cos(2 \phi) + b \sin(2 \phi) \right) r^2  + w_2r\sin(\phi) + w_1 r\cos(\phi) \right)}.\label{conmg2}
\end{align}
}

The Cartesian form of the magnetic field and the {electrostatic} potential reads
\begin{eqnarray}
&&
\vec{B}(\vec{x})=(- { b_\phi}y, { b_\phi}x,{ b_Z}),\label{wcs2}
\\ \nonumber
&&  W(\vec{x})  = { b_\phi \left(  -\frac{1}{8}   b_\phi(x^2+y^2)^2   +  w_3 (x^2+y^2) -{b_Z} \left( \frac{a}{2} \left( x^2 - y^2 \right)+ b x y \right) +  w_1 x + w_2 y \right)}.
\end{eqnarray}
The corresponding vector potential can be chosen as
\begin{eqnarray}
\vec{A}(\vec{x})= \left(-\frac{1}{2}{ b_Z}y,  \frac{1}{2}{ b_Z}x,-\frac{1}{2}{ b_\phi}(x^2+y^2)\right).\label{aavec1}
\end{eqnarray} 
The coefficients of the integrals (\ref{Y1}) and (\ref{Y2}) we can derive from (\ref{src}), (\ref{Src}) and the solution of (\ref{hy1m}) and (\ref{hy2m})
\begin{eqnarray}
&& m(r, \phi, Z) =-\frac{1}{2} \Big({ b^2_\phi} r^2 +2 { b^2_Z} -4 { b_\phi}{  w_3} + 2{ b_\phi} { b_Z}a\Big)a r^2\cos^2(\phi) \nonumber\\&&\qquad - \frac{1}{2}( { b^2_\phi}r^2   +2{ b^2_Z} -4 { b_\phi}{ w_3}+2{ b_\phi}{ b_Z} a)br^2\cos(\phi)\sin(\phi)\nonumber\\&&\qquad +(2{ b_\phi}{ w_1}a + { b_\phi}{ w_2}b + { w_1} { b_Z})r\cos(\phi)+({ b_\phi}{ w_1}b + { w_2} { b_Z})r\sin(\phi) \nonumber\\&&\qquad -\frac{1}{4}\Big( { b_\phi} { b_Z}r^4 + 2({ b_\phi}{ b_Z} b^2 -{ b^2_Z}a  - { b_\phi}s_{Z3} - { 2 w_3} { b_Z})r^2\Big),  \label{ccMg1}
 \\&& M(r, \phi, Z) =\frac{1}{4} { b^2_\phi}r^4 +\frac{1}{2}{ b_\phi}S_{Z1}r^2   \label{ccmg1}  
 \end{eqnarray}
 (again unique up to irrelevant additive constants).
This system is a generalization of the system which was presented as an example in the paper \cite{Sno1} for $b=0$.
{ From the presence of the arbitrary parameters $s_{Z3}$, $S_{Z1}$ in the integrals which are not in the magnetic field nor in the potential one easily infers that the system} allows { a} first order integral
\begin{eqnarray}
X_3^{1}=p^A_3+\frac{1}{2} { b_\phi}(x^2+y^2),\label{fst2}
\end{eqnarray} 
which can be interpreted as a square root of the integral $X_2$ since we have the relation
\begin{eqnarray}
X_2 =(X_3^{1})^2+S_{Z1}X_3^{1}.
\end{eqnarray}
In the choice of gauge (\ref{aavec1}) it reduces to ${ X_3^1}=p_3$.
Thus $X_1$ and $X^1_3$ form a commuting pair of integrals equivalent to $X_1$ and $X_2$.

The system characterized by the magnetic field~\eqref{conmg1} and the electrostatic potential~\eqref{conmg2} is for generic values of its parameters only integrable\footnote{As long as integrals at most quadratic in the momenta are {considered. Search for hypothetical additional} higher order integrals seems presently computationally unachievable.} as we verified by an explicit, if rather lengthy, computation assuming the hypothetical additional integral of the form (\ref{X4}). However, {we obtain additional integrals for particular values of the parameters}, as we review case by case below.

\subsubsection{Superintegrability with ${b_Z}=0$}
In this case, the magnetic field and the potential read
\begin{eqnarray}
&&B_r(r, \phi, Z) = 0,\quad B_\phi(r, \phi, Z) = {b_\phi}r,\quad  B_Z(r, \phi, Z) = 0,\label{mgiva}
\nonumber\\&&
W(r, \phi, Z) = -\frac{1}{8}{b^2_\phi}r^4 + {b_\phi} \left( {w_3}r^2+ {w_1}r\cos(\phi) + {w_2}r\sin(\phi) \right).\label{potential3a}
\end{eqnarray}
The Cartesian form of them are given by 
\begin{eqnarray}
&&\vec{B}(\vec{x})=(- {b_\phi}y, {b_\phi}x,0),\nonumber
\\&& W(\vec{x}) =-\frac{1}{8}{b^2_\phi}(x^2+y^2)^2 + {b_\phi} \left( {w_3}(x^2+y^2) + {w_1}x + {w_2}y \right).  \label{elec1}
\end{eqnarray}
It is seen that the system described by (\ref{elec1}) is a particular form of the Case I(d) in \cite{Mar5} with the parameter identification $a_1=-\frac{1}{2}{b_\phi}$,  $b_1={b_\phi} {w_3}$, $b_2={b_\phi}{w_1}$, $b_3={b_\phi}{w_2}$, $a_2=a_3=0$ therein. 
In this case, the system allows one first order integral, namely (\ref{fst2}), which can again be considered instead of $X_2$.
The parameters $a,b$ are absent in the magnetic field and the potential (\ref{potential3a}), thus we may split $X_1$ (\ref{Y1}) into three parts according to the coefficients of $a$, $b$ and $a, b$ free terms as follows,
\begin{eqnarray}
&& X_1^{a}=\left(\cos(\phi)p^A_r-\frac{\sin(\phi)}{r}p^A_\phi \right)^2- {b_\phi}r^2\cos^2(\phi)p_Z^A -\frac{1}{2} {b_\phi}({b_\phi}r^2  -4 { w_3})r^2\cos^2(\phi) \nonumber\\&&\qquad\quad+ 2{b_\phi}{w_1}r\cos(\phi),\nonumber
\\
&& X_1^{b}=\left(\cos(\varphi) p^A_r - \frac{\sin(\varphi)}{r}p^A_\varphi \right)\left(\sin(\varphi) p^A_r + \frac{\cos(\varphi)}{r}p^A_\varphi \right) -{b_\phi}r^2\sin(\phi)\cos(\phi)p_Z^A \nonumber\\&&\qquad\quad -\frac{1}{2} {b_\phi}({b_\phi}r^2 -4{w_3})r^2\sin(\phi)\cos(\phi) + {b_\phi}{w_2}r\cos(\phi) + {b_\phi}{w_1}r\sin(\phi),
\\
&& X_1^{c}= p^A_\phi p^A_Z+({w_2}\cos(\phi) - {w_1}\sin(\phi))p^A_r+ \left(\frac{{b_\phi}}{2} r^2 -\frac{1}{r} {w_2}\sin(\phi) -\frac{1}{r} {w_1}\cos(\phi)\right. \nonumber\\&&\qquad\qquad \left.- {2 w_3}\right)p_\phi^A + s_{Z3}p_Z^A+\frac{1}{2}{b_\phi}s_{Z3}  r^2,   \nonumber
\end{eqnarray}
which must independently be integrals of the system (\ref{potential3a}).
The Cartesian form of the integrals reads
\begin{eqnarray}
&& X_1^{a}=(p_1^A)^2- {b_\phi}x^2p_3^A -\frac{1}{2} {b^2_\phi}x^2(x^2+y^2) +  2 {b_\phi}{w_3}x^2 + 2{b_\phi}{w_1}x,\nonumber
\\
&& X_1^{b}=p_1^A p_2^A -{b_\phi}xyp_3^A  -\frac{1}{2} {b^2_\phi}xy(x^2+y^2) +2 {b_\phi}{w_3}xy + {b_\phi}{w_2}x + {b_\phi}{w_1}y,\label{x1b1}
\\
&& X_1^{c}= l^A_3 p^A_3+{w_2}p^A_1 - {w_1}p^A_2+ \Big(\frac{1}{2}{b_\phi}(x^2+y^2) - {2 w_3}\Big)l_3^A + s_{Z3}p_3^A \nonumber \\&&\qquad\qquad +\frac{1}{2}{b_\phi}s_{Z3}(x^2+y^2).\nonumber
\end{eqnarray}
The sets of integrals $\{H, X_1^{a}, X_1^{b},  X_3^1\}$ or  $\{H, X_1^{a},  X_1^{c}, X_3^1\}$ are functionally independent. Thus the Hamiltonian system given by~\eqref{elec1} is minimally superintegrable. Searching for additional at most quadratic integrals, we found that there is no other independent integral. The integrals $H, X_1^{a}, X_1^{b}, X_1^{c}, X_3^1$ of the system close to form the following Poisson algebra,
\begin{eqnarray}
&& \{X_1^a,X_1^b\}_{P.B.} = 2{b_\phi}(s_{Z3}X_3^1-X_1^c ),\nonumber
\\&&
 \{X_1^a,X_1^c\}_{P.B.} = - 2 (X^1_3  -2  { w_3})X_1^b+2{b_\phi}{w_1}{w_2},
 \\&&
 \{X_1^b,X_1^c\}_{P.B.}=(X_3^1-{2 w_3})((X_3^1)^2+2X_1^a-2H)+{b_\phi}({w^2_2}-{w_1^2}),\nonumber
 \\&& \{X_1^a,X_3^1\}_{P.B.} = \{X_1^b,X_3^1\}_{P.B.}=\{X_1^c,X_3^1\}_{P.B.}=0.\nonumber
\end{eqnarray} 
The system~\eqref{elec1} separates in the Cartesian coordinates (but not in the cylindrical coordinates) as is easily verified by the Levi--Civita condition \cite{levi04}.

\subsubsection{Superintegrability with ${b_Z}=0$,  ${w_1} = 0$, ${w_2} = 0$}
In this case, the magnetic field and the electrostatic potential read
\begin{eqnarray}
&&B_r(r, \phi, Z) = 0,\quad B_\phi(r, \phi, Z) = {b_\phi}r,\quad  B_Z(r, \phi, Z) = 0,
\nonumber\\&&
W(r, \phi, Z) = -\frac{1}{8} {b^2_\phi}r^4 +{b_\phi}{ w_3} r^2.\label{pp1}
\end{eqnarray}
The Cartesian form of the magnetic field and the electrostatic potential are given by 
\begin{eqnarray}
\vec{B}(\vec{x})=(-{b_\phi}y, {b_\phi}x, 0), \quad W(\vec{x}) = -\frac{1}{8} {b^2_\phi}(x^2+y^2)^2 +{b_\phi}{w_3}(x^2+y^2).\label{cw1}
\end{eqnarray}
The system described by (\ref{pp1}) is a particular form of the Case I(d) in \cite{Mar5} with $a_1=-\frac{1}{2}{b_\phi}$, $a_2=a_3=0$, $b_1={b_\phi} { w_3}$, $b_2=b_3=0$
for which also invariance under rotations is present, see integral $X_3^2$ below. The scalar coefficients of the second-order integrals (\ref{Y1}) and (\ref{Y2}) read (\ref{ccmg1})-(\ref{ccMg1}) with ${b_Z} = 0$,  ${w_1} = 0$, ${w_2} = 0$. 
As the real constants $a$ and $b$ are again absent in the magnetic field and potential (\ref{pp1}), we can split the second-order integral $X_1$ (\ref{Y1}) into three parts
\begin{eqnarray}
\nonumber && X_1^{a}=\left(\cos(\phi)p^A_r-\frac{\sin(\phi)}{r}p^A_\phi \right)^2- {b_\phi} r^2\cos^2(\phi)p_Z^A -\frac{1}{2} {b_\phi}({b_\phi} r^2 -4{w_3}) r^2 \cos^2(\phi),
\\
&& X_1^{b}=\left(\cos(\varphi) p^A_r - \frac{\sin(\varphi)}{r}p^A_\varphi \right)\left(\sin(\varphi) p^A_r + \frac{\cos(\varphi)}{r}p^A_\varphi \right) -{b_\phi} r^2\sin(\phi)\cos(\phi)p_Z^A \nonumber\\&&\qquad\qquad  - \frac{1}{2} {b_\phi}({b_\phi} r^2 -4 { w_3}) r^2\sin(\phi)\cos(\phi),
\\
\nonumber && X_1^{c}= p^A_\phi p^A_Z+ \left(\frac{1}{2} {b_\phi}r^2 - {2 w_3}\right)p_\phi^A + s_{Z3} p_Z^A +\frac{1}{2} {b_\phi}s_{Z3} r^2.
\end{eqnarray}
They are independently integrals of our Hamiltonian system (\ref{pp1}), i.e., 
\begin{eqnarray}
\{H, X_1^{a}\}=0,\quad \{H, X_1^{b}\}=0,\quad \{H, X_1^{c}\}=0.
\end{eqnarray}
In the Cartesian coordinates, they become
\begin{eqnarray}
\nonumber X_1^{a} & = & (p_1^A)^2- {b_\phi} x^2p_3^A  -\frac{1}{2} {b^2_\phi} x^2(x^2+y^2)+2 {b_\phi}{w_3} x^2,
\\
 X_1^{b} & = & p_1^A p_2^A -{b_\phi} x y p_3^A - \frac{1}{2} {b^2_\phi}xy(x^2+y^2) +2 {b_\phi}{w_3} x y,
\\
\nonumber  X_1^{c} & = & l^A_3 p^A_3+ \left(\frac{1}{2} {b_\phi}(x^2+y^2) - {2 w_3}\right)l_3^A + s_{Z3} p_3^A +\frac{1}{2}{b_\phi}s_{Z3}(x^2+y^2).
\end{eqnarray}
The system described by (\ref{pp1}) allows the first--order integral (\ref{fst2}) and 
\begin{eqnarray}
 X_3^{2}= p^A_\phi =l^A_3,\label{fsti}
\end{eqnarray}
which are in involution with the Hamiltonian and mutually with each other, that is, 
\begin{eqnarray}
\{H,X_3^{1}\}_{P.B.}=0=\{H,X_3^{2}\}_{P.B.},\quad \{X_3^1,X_3^{2}\}_{P.B.}=0,
\end{eqnarray}
thus the system becomes minimally superintegrable of cylindrical type with first order integrals \cite{Mar3}.
The first order integral $X_3^{1}$ plays the role of a central element of our system, that is,
\begin{eqnarray}
\{X_3^{1},X_1^{a}\}_{P.B.}=\{X_3^{1},X_1^{b}\}_{P.B.}= \{X_3^{1},X_1^{c}\}_{P.B.}=0
\end{eqnarray}
and $X_3^{2}$ satisfies the following Poisson bracket relations with the second order integrals of motion
\begin{eqnarray}
 \{X_3^{2},X_1^{c}\}_{P.B.}=0,\quad
\{X_3^{2},X_1^{a}\}_{P.B.}=2 X_1^{b},\quad
 \{X_3^{2},X_1^{b}\}_{P.B.}=-((X_3^1)^2-2H) - 2X_1^a.
\end{eqnarray}
The integral $X_1^{c}$ can be expressed in terms of the first-order integrals $X_3^{1}$ and $X_3^{2}$ as
\begin{eqnarray}
X_1^{c}=X_3^{1}X_3^{2}-{2 w_3} X_3^{2}+s_{Z3}X_3^{1}.
\end{eqnarray}
In addition, we may express square of the integral $X_1^{b}$ as a polynomial expression in the other integrals in the following way,
\begin{eqnarray}
 (X_1^{b})^2=(2H-(X_3^1)^2-X_1^a)X^a_1 +{b_\phi}(X^1_3-{2 w_3})(X^2_3)^2,
\end{eqnarray}
which indicates that $X_1^{b}$ is linearly independent but functionally dependent. The polynomial algebra of our integrals closes with the last relevant Poisson bracket
\begin{eqnarray}
\{X_1^a,X_1^b\}_{P.B.} = 2{b_\phi}(s_{Z3} X_3^1-X_1^c).
\end{eqnarray}
Solving the determining equations for an hypothetical additional second order integral (\ref{X4}) we find that no other functionally independent second order integral exists, thus at the level of second order integrals the system (\ref{pp1}) is minimally superintegrable.
The system (\ref{pp1}) separates both in the Cartesian and cylindrical coordinates \cite{levi04}.

\section{The systems and integrals of motion for $a=b=0$}\label{sec5}
In this section, we conclude our computation by considering the parameters $a=b=0$ in the second-order integral (\ref{Y1}). Similarly as above, we solve the leading order determining equations coming from
\begin{eqnarray}
\{H,X_1\}_{P.B.}=\{H,X_2\}_{P.B.}=\{X_1,X_2\}_{P.B.}=0. \label{pab1}
\end{eqnarray}
In this case, some of the equations in (\ref{y1y2s}) vanish identically. The leading order equations (\ref{hy1s}), (\ref{hy1S}) and (\ref{y1y2s}) in this case imply
\begin{eqnarray}
 B_r(r, \phi, Z) = 0,\quad B_\phi(r, \phi, Z) = B_\phi(r),\quad  B_Z(r, \phi, Z) = B_Z(r),
\end{eqnarray}
where $B_\phi(r)$ and $B_Z(r)$ are arbitrary functions together with the corresponding structure of the functions $s_\lambda(r,\phi,Z)$ and $S_\lambda(r,\phi,Z)$ depending on these functions and {a} certain number of integration constants. However, solving the lower order equations (\ref{hy1m}), (\ref{hy1w}), (\ref{hy2m}), (\ref{hy2w}), assuming that the system does not possess two commuting first order integrals of the form $l^A_Z+\dots$ and $p^A_Z+\dots$ (such integrals then trivially allow to construct integrals of the form (\ref{Y1}) and (\ref{Y2}) with $a=b=0$; however, the corresponding systems would trivially be of {the} standard cylindrical type, {cf.} \cite{Mar1} and involve arbitrary functions of $r$, see equation (76) therein) and has {a} nonvanishing magnetic field, we arrive at the magnetic field 
\begin{eqnarray}
 B_r(r, \phi, Z) = 0,\quad B_\phi(r, \phi, Z) = S_{Z4}r,\quad  B_Z(r, \phi, Z) = - s_{Z1}r,\label{mga1}
\end{eqnarray}
which is of the same form as in (\ref{fmg1}).
The electrostatic potential in this case takes the form
\begin{eqnarray}
 W(r, \phi, Z) =  -\frac{1}{8}S_{Z4}^2r^4 -\frac{1}{2} S_{Z4}s_{\phi 1} r^2 + S_{Z4}s_{r2}r\sin(\phi) - S_{Z4} s_{r1} r\cos(\phi).\label{wma1}
\end{eqnarray}
We observe that while the equations in the case $a=b=0$ are somewhat different and details of the computation differ, their solution can be obtained as a special subclass of (\ref{fmg1}) and (\ref{fpw1}) setting $a=b=0$ after suitable relabelling of the parameters (i.e., integration constants).

\section{Conclusion}\label{sec6}
The main result of this paper is the classification of nonstandard integrable systems with magnetic fields whose integrals have the form
\begin{eqnarray}
X_1= l^A_3p_3^A+a(p^A_1)^2+bp^A_1p_2^A+\dots,\quad
X_2=(p^A_3)^2+\dots, \quad a,b\in \mathbb{R}.
\end{eqnarray}
The choice of this class {among the nonstandard classes of Theorem~\ref{EnvAlg_MainTheor}} was dictated by computational feasibility since it was expected that the equations can be analysed in the cylindrical coordinates. We have seen that under the assumption that the integrals do not reduce to the standard form characterizing the Cartesian or cylindrical type integrals there is a unique family of such systems characterized by the magnetic field~{\eqref{conmg1}, namely} 
\begin{eqnarray*}
B_r(r, \phi, Z) = 0,\quad B_\phi(r, \phi, Z) = b_{\phi}r,\quad  B_Z(r, \phi, Z) = b_Z r
\end{eqnarray*}
and the electrostatic potential~{\eqref{conmg2}, namely}
\begin{align*}
&&W(r, \phi, Z) = {b_{\phi} \left( -\frac{1}{8} b_{\phi}r^4  + w_3 r^2 - \frac{b_Z }{2} \left( a \cos(2 \phi) + b \sin(2 \phi) \right) r^2  + w_2r\sin(\phi) + w_1 r\cos(\phi) \right)},
\end{align*}
{(for the equivalent expressions in Cartesian coordinates see Theorem~\ref{NewTheorem}).} When $b=0$, the system (\ref{conmg1})- (\ref{conmg2}) becomes the system (42)--(43) of \cite{Sno1}. Thus this system generalizes the system (42)--(43) of \cite{Sno1} to nonvanishing values of $b$.
We have checked that among the nonstandard systems {of} this form none possesses additional independent integrals of motion. However, in suitable limits of the parameters the system enhances its symmetries, falls {also} into one or two standard classes and {possesses} additional integrals making it superintegrable.

The results presented here complement the results of \cite{Kub2022}. Therein classes from Theorem~\ref{EnvAlg_MainTheor} were considered which generalize {the} structure of integrals characterizing physically relevant cases of cylindrical and spherical separability. In our present paper, we have focused on a class that does not have a direct counterpart in the absence of {a} magnetic field. We notice that the resulting system (\ref{conmg1}) and (\ref{conmg2})
in {the} limit of vanishing magnetic field $b_{\phi}=b_{Z}=0$ becomes simply the free Hamiltonian, separating in any orthogonal coordinate system. If only {$b_{\phi}=0$}, we have the system with {a} constant magnetic field and vanishing electrostatic potential. If {$b_\phi=0$}, we have {a} system which falls into the standard Cartesian class and is superintegrable.

One of {the} observations coming from the results presented here and in \cite{Kub2022} is that for almost all nonstandard type systems discovered so far ({except} the system (3.48) in \cite{Kub2022}), i.e., those not reducing to the leading order structure of its integrals corresponding to {an} orthogonal separation, one of the integrals reduces to a first order one. To better understand the reason behind this provides motivation to analyse the remaining classes in Theorem~\ref{EnvAlg_MainTheor} despite the fact that for most of them it is so far not clear in which coordinate system they may be tractable.

\section*{Acknowledgement}
FH was supported by the project grant CZ.02.2.69/0.0/0.0/18\_053/0016980 Mobility CTU - STA, Ministry of Education, Youth and Sports of the Czech Republic, co-financed by the European Union. L\v{S} was supported by the project of the Ministry of Education, Youth and Sports of the Czech Republic CZ.02.1.01/0.0 /0.0/16\_019/0000778 Centre of Advanced Applied Sciences, co-financed by the European Union.

\end{document}